\documentclass[%
 reprint,
showpacs,
 amsmath,amssymb,
 aps,
pra,
]{revtex4-1}

\usepackage{graphicx}
\usepackage{dcolumn}
\usepackage{bm}
\usepackage{url}
\usepackage{color}
\usepackage{epstopdf}
 \usepackage{ifsym}
\usepackage[caption=false]{subfig}
\newcommand\crule[3][black]{\textcolor{#1}{\rule{#2}{#3}}}
\begin{document}


\title{Pulse chirp increasing pulse compression followed by positive resonant radiation in fibers}

\author{Joanna McLenaghan}
\author{Friedrich K\"onig}%
 \email{fewk@st-andrews.ac.uk}
\affiliation{%
 School of Physics and Astronomy, SUPA, University
of St. Andrews, North Haugh, St. Andrews, KY16 9SS, UK\\
 \\
}%

\date{\today}

\begin{abstract}
Pulse self-compression followed by the generation of resonant radiation is a well known phenomenon in non-linear optics. Resonant radiation is important as it allows for efficient and tunable wavelength conversion. We vary the chirp of the initial pulse and find in simulations and experiments that a small positive chirp enhances the pulse compression and strongly increases the generation of  resonant radiation. This result corroborates previously published simulation results indicating an improved degree of pulse compression for a small positive chirp \cite{McLenaghan:2014}. It also demonstrates how pulse evolution can be studied without cutting back the fiber. 

\begin{description}
\item[PACS numbers]
\end{description}
\end{abstract}

\pacs{Valid PACS appear here}
\maketitle


\section{Introduction}

Resonant radiation (RR, optical Cherenkov radiation, non-dispersive radiation) arises when an optical soliton is perturbed by higher order dispersion and transfers energy to radiation in the normal dispersion region of the spectrum \cite{Akhmediev:1995hc}. It has also been observed for other pump pulses, for example 3-dimensional X-waves in bulk media \cite{Rubino:2012ly}. 

RR has potential practical applications as well as theoretical implications. For applications, RR is a source of visible light for use in the field of photobiology \cite{Tu:2013}, to produce broadband visible wavelength astrocombs \cite{Chang:12}, and to cancel the Raman shift of the driving pulse by spectral recoil \cite{Skryabin:2003qy}. Recent research into the process of RR generation has described it as resulting from the scattering of pulse photons at the optical event horizon \cite{Choudhary:12}. This links RR to similar mode conversion processes such as the generation of the optical analogue of Hawking radiation \cite{Belgiorno:10}. At the same time the scattering of light also creates negative resonant radiation (NRR) \cite{Rubino:2012ly}. This involves a coupling between positive and negative frequencies and therefore is a novel amplification process in optics \cite{Rubino:2012fk}.

Progress has been made to meet the requirements for a practical RR based light source. Typical applications of RR require narrow-band tunable light with powers in the mW range and high stability. Tunability is achieved either by using tapered fibers \cite{Lu:05,Lu:06}, by varying the input pulse polarization \cite{Mitrofanov:06,Ivanov:06} or by tuning the pump wavelength \cite{Tu:09}. The RR bandwidth is reduced by tuning the input pulse wavelength away from the fiber zero dispersion wavelength \cite{Tu:09}. The pulse to pulse energy stability is improved by using an all fiber set up \cite{Liu:12,Liu:13} or tapered fibers \cite{Lu:06}. In order to reach mW power levels, a high conversion efficiency between the pump pulse and the RR is required. Various groups have studied the efficiency dependence on: pump wavelength separation from the fiber ZDW, pump pulse length, pump power, fiber core size and fiber third order dispersion \cite{Chang:10,Roy:2009kl,Tu:09b,Tu:09}.
The generation of the RR depends strongly on the overlap of the driving pulse spectrum with the phase matched RR wavelength. Therefore the strongest RR signal will be produced when the input pulse compresses most efficiently and its spectrum expands to the greatest extent. 
A crucial parameter here is the input pulse frequency chirp. This can be used to delay the nonlinear pulse evolution by introducing a linear chirp which must first be compensated. This effect has been studied in relation to supercontinuum generation \cite{Cheng2011,Fu2004,Zhang2007,Zhu2004} and in particular spectral broadening \cite{Tianprateep2005, Tianprateep2004}. In the latter two papers numerical and experimental results show this delayed pulse compression and demonstrate that the greatest spectral broadening in the output of a fiber can be expected when the input pulse chirp is positive and equal to the magnitude of the negative dispersion over the whole fiber length. 
Delayed compression due to chirp is therefore an established idea but one might expect the chirp to be eventually compensated and thus to have no impact on the degree of pulse compression. However, in this paper we show that the chirp can be used to increase the degree of compression and to control the emisson of resonant radiation. 
In a previous paper \cite{McLenaghan:2014} we presented the first detailed investigation of the effect of pulse chirp beyond the delayed nonlinear compression. In that paper we showed numerically and experimentally that, for shorter pulses and fibers than had been considered previously, interesting effects arise such as an increased degree of pulse compression for a small positive chirp compared to zero chirp. We therefore expect to be able to observe an increased RR generation efficiency for this optimum chirp. 
In this paper we present experimental results testing this behaviour by measuring the output RR from a fiber as a function of the input chirp. 
Additionally, we apply the recent technique described in \cite{McLenaghan:2014}, varying the chirp to change the distance propagated by the RR between its generation and the fiber end. In this way we can qualitatively investigate its evolution without the need to cut the fiber. 

In the next section we discuss the phase matching condition for RR and how we can use this with the fiber dispersion relations to predict the RR wavelength. In the following section the propagation of chirped pulses is considered, in particular the effect of pulse compression. Simulation results are then presented showing how the chirp affects pulse compression.  We find that chirp can be used to move the position of pulse compression and that contrary to a naive understanding the chirp can also be used to vary the compression ratio. Finally experimental results are presented to demonstrate how this variation in the compression ratio affects the generation of RR. 

\section{Phase matching of resonant radiation}

Light propagation in the fiber is governed by the generalised nonlinear Schr\"{o}dinger equation (GNLSE) \cite{Agrawal:01}.  When higher order and absorption terms are neglected the GNLSE can be simplified to the nonlinear Schr\"{o}dinger equation (NLSE) of which the family of fundamental and higher order solitons are solutions. These are characterised by a soliton order $\rm{N}^{2}=\gamma\rm{P}_{0}\rm{T}_{0}^{2}/\left|\beta_{2}\right|$, where $\rm{P}_{0}$ and $\rm{T}_{0}$ are the soliton peak power and length ($\tau=1.763\,\rm{T}_{0} $ is the FWHM length), $\gamma$ is the non-linearity and $\beta_{2}$ the second order dispersion parameter of the fiber \cite{Agrawal:01}. The fundamental soliton is characterised by $\rm{N}=1$ and the higher order solitons by larger integers.  With small higher order dispersion terms (third order and up) included the soliton is perturbed and will lose energy to dispersive RR and NRR in the normal group velocity dispersion region. Higher order solitons become unstable and fission into $\rm{N}$ pulses occurs, each of which will lose energy to RR and NRR \cite{Herrmann:2002fj,Husakou:2001rt}. 
In order for a RR signal to build up in the fiber the RR waves generated at different times must interfere constructively. This occurs when the following momentum conserving phase matching condition is satisfied \cite{Cristiani:04}:

\begin{equation}
\sum_{n\geq2}\frac{\beta_{n}}{n!}\left(\omega_{\mathrm{RR}}-\omega_{\mathrm{P}}\right)^{n}=\frac{\left(2N-1\right)^{2}\gamma P_{0}}{2N^{2}}\label{eq:PM0}
\end{equation}

Where $\omega_{\mathrm{RR}}$ and $\omega_{\mathrm{P}}$ are the RR and pulse angular
frequencies, respectively. We express (\ref{eq:PM0})  as a conservation of frequency in the frame comoving with the pulse \cite{Rubino:2012ly,Philbin:2008fr}:

\begin{equation}
\omega'_{\mathrm{P}}=\omega'_{\mathrm{RR}}\label{eq:PM}
\end{equation}

Here  $\omega' = \omega-v\, k$ is the Doppler shifted frequency in the comoving frame of the pulse of velocity $v$. The mode propagation constant is related to the fiber refractive index by $\beta(\omega)=\frac{n\left(\omega\right)\omega}{c}$.  

\begin{figure}
\begin{centering}
\includegraphics[scale=0.3]{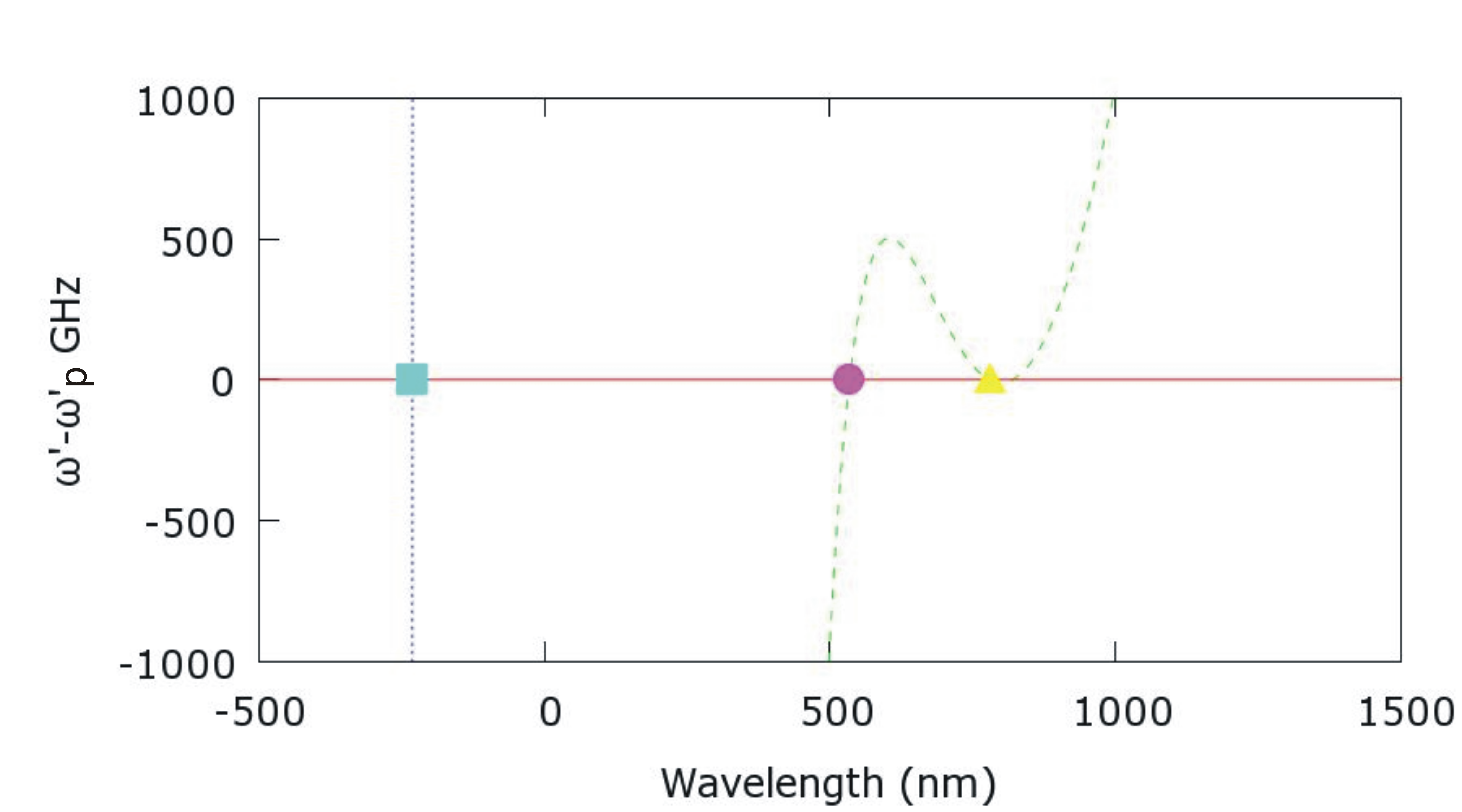}
\par\end{centering}
\begin{centering}
\caption{Co-moving frame frequency $\omega'$ as a function of wavelength for
a pulse at $800$\,nm. Markers indicate the pulse wavelength ( \textcolor{yellow}{$\blacktriangle$} ), the RR in the visible ( {\large \textcolor{magenta}{$\bullet$}}), and the NRR in the UV({\small \crule[cyan]{0.2cm}{0.2cm} }).\label{fig:1}}
\par\end{centering}
\end{figure}

Formulation (\ref{eq:PM}) of the phase matching condition makes it simple to find solutions using graphical methods. Figure \ref{fig:1} shows $\omega'$ around $\omega'_{\mathrm{P}}$ as a function of the lab frame wavelength for fiber 1  (NL-1.5-590 by NKT Photonics) used in our experiment. We assume a pulse at 800$\,\mathrm{nm}$ and negligible nonlinear contributions from the input pulse. Hence, Fig.  \ref{fig:1} shows the RR as a solution to the phasematching condition in the visible. The NRR solution can also be seen at `negative wavelengths'.

\section{Propagation of chirped pulses}

The evolution of a pulse in a fiber will be influenced mainly by self-phase modulation (SPM) and group velocity dispersion (GVD). The former affects the pulse spectrum and the latter changes the relative velocity of the pulse frequencies. The interplay of these two effects allows stable solitons to form in the absence of additional higher order effects. The relative strength of SPM and GVD determines the formation of a fundamental or a higher order soliton. Higher order solitons initially compress in time and broaden spectrally due to GVD and SPM. 
When a pulse generates RR, the generation is driven by the amplitude of the pulse at the RR frequency; hence the greater the degree of spectral broadening the more RR will be generated. 

An initially chirped pulse has the same spectrum as an unchirped pulse, but the instantaneous frequency will vary across the pulse, broadening the pulse temporally.
An anomalous fiber dispersion will compensate an initial positive chirp and thus the pulse will compress towards its unchirped length. The interplay of SPM and GVD will then compress the pulse further as for unchirped input pulses until a minimum pulse length is reached at the `compression point'. Therefore, a variable initial chirp can be used to compress the pulse at a controlled fiber length. Hence, the strongest RR production is expected at the compression point with the amount critically dependent on the compression ratio.

As we will see in the simulations and experiments, the compression ratio will be greater for a small positive chirp compared to zero chirp and then reduce as the chirp becomes larger. 

\section{Simulations of chirped pulse propagation}

We perform numerical simulations of pulse propagation under the GNLSE.  We use a standard split-step Fourier tool \cite{Paschotta}, which takes into account the full dispersion profile as well as nonlinear and Raman effects. We use dispersion profiles determined from RR measurements for the two fibers used in the experiments \cite{Rubino:2012ly}. Fiber 1 (NL-1.5-590, NKT Photonics, Ltd.) has a zero-dispersion wavelength (ZDW) of 685$\,$nm and fiber 2 (NL-1.6-615, NKT Photonics, Ltd.) a ZDW of 697$\,$nm. Both fibers have anomalous dispersion at longer wavelengths. Propagations of hyperbolic secant pulses with varying linear chirp are simulated. 

The main factors for the generation of RR are the evolution of the pulse spectrum and the pulse peak power, stimulating and driving RR production, respectively.  The simulation parameters are matched to the experimental parameters. We used a $12\,\mathrm{fs}$ hyperbolic secant pulse with a power corresponding to a soliton order of $N=2.25$ and a center wavelength of $800\,\mathrm{nm}$ in the anomalous dispersion region. We also extended the investigation to different soliton orders and pulse lengths. 

\begin{figure}
\begin{centering}
\begin{tabular}{cc}
\subfloat[ \label{fig:2a}]
{\begin{centering}
\includegraphics[scale=0.16]{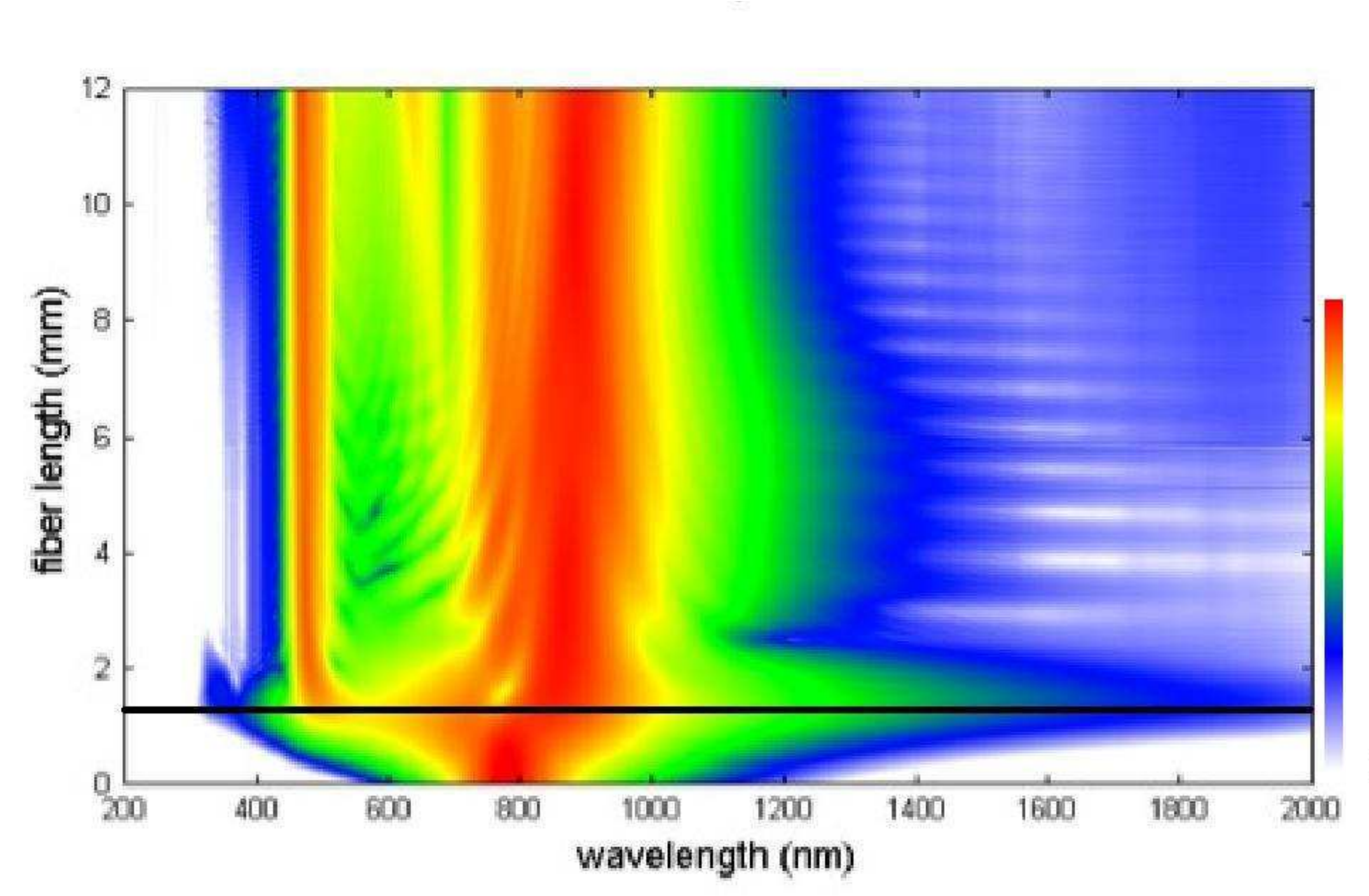}  
\end{centering}
}
&
\hspace{-4mm}
\subfloat[ \label{fig:2b}]
{\begin{centering}
 \includegraphics[scale=0.16]{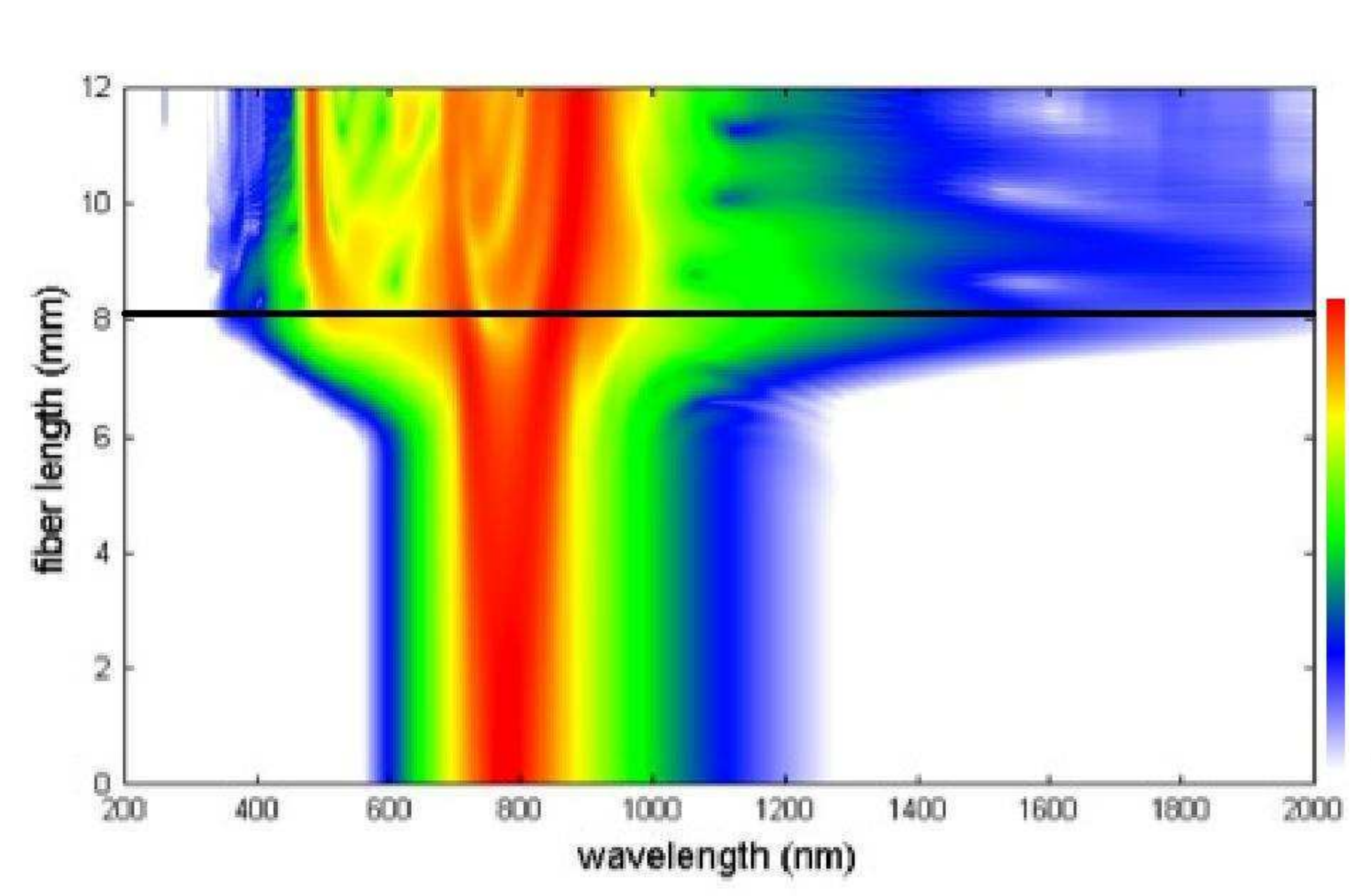}
\end{centering}
}
 \tabularnewline
\end{tabular}
\par\end{centering}
\caption{Pulse spectral evolution during propagation
for two different input pulse chirps exemplified with fiber 1 ((a): $0\:\mathrm{fs^{2}}$; (b): $305\:\mathrm{fs^{2}}$). $N$=$2.25$, $12\,$fs hyperbolic secant pulses were used. The horizontal line indicates the maximum spectral broadening. Figure taken from \cite{McLenaghan:2014}.
\label{fig:2}}
\end{figure}

\begin{figure}
\begin{centering}
\includegraphics[scale=0.3]{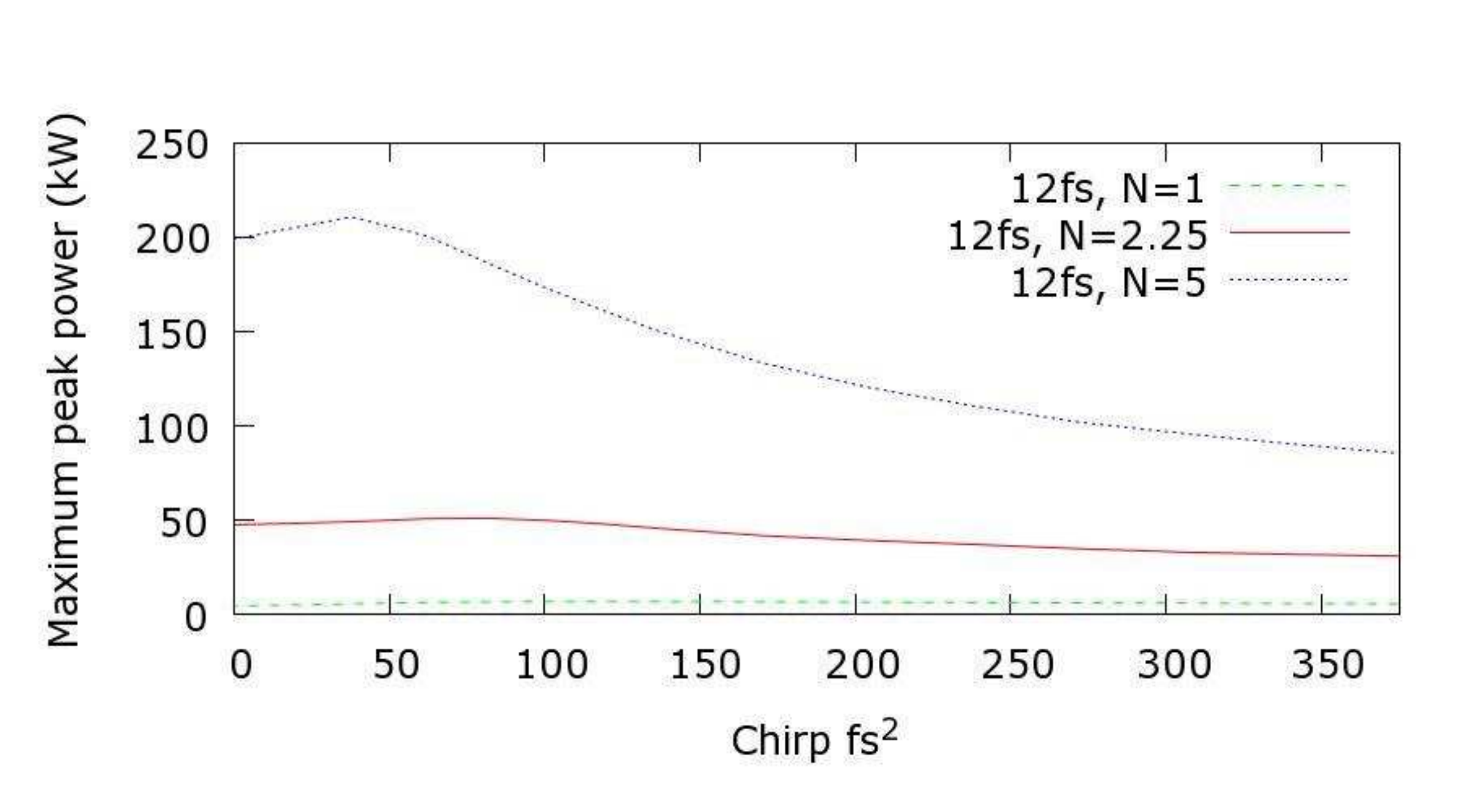}
\par\end{centering}
\caption{Peak power of compressed pulse in fiber 1 as a function of pre-chirp applied to the $12\,$fs input pulse. \label{fig:3} }
\end{figure}

Detailed results of the fiber 1 simulations can be found in \cite{McLenaghan:2014}. Here we briefly review the results and conclusions relevant to RR generation. Two example spectral evolutions for this fiber are shown in figure \ref{fig:2} for an unchirped (a) and a chirped pulse (b). For zero input chirp the input pulse spectrum expands to almost three octaves within the first $2\,\mathrm{mm}$ of fiber and then contracts again rapidly. RR is generated when the spectrum expands and reaches the RR wavelength between $400$ and $500\,\mathrm{nm}$. Similar results are found for fiber 2. 
For the large input chirp in Fig. \ref{fig:2} (b) the behavior is similar, however, the point of temporal compression and spectral expansion is considerably further along the fiber. In addition the extent of the spectral broadening has slightly decreased. 

As the pulse compresses and broadens, its peak power will rise to a maximum and then decrease. As we see, the RR is alomst entirely generated where the pulse compresses, i.e. spectrally broadens, and assumes the maximum peak power. Hence it is instructive to find out how efficient the pulse compression is as the pulse chirp is varied. Figure \ref{fig:3} shows the maximum, compressed peak power as a function of the input pulse chirp. The peak power assumes a maximum for small positive chirps exceeding the power for zero chirp. For larger positive chirps the maximum peak power decreases, indicating that the pulse is not compressing as much. 

\begin{figure}
\begin{centering}
\includegraphics[scale=0.50]{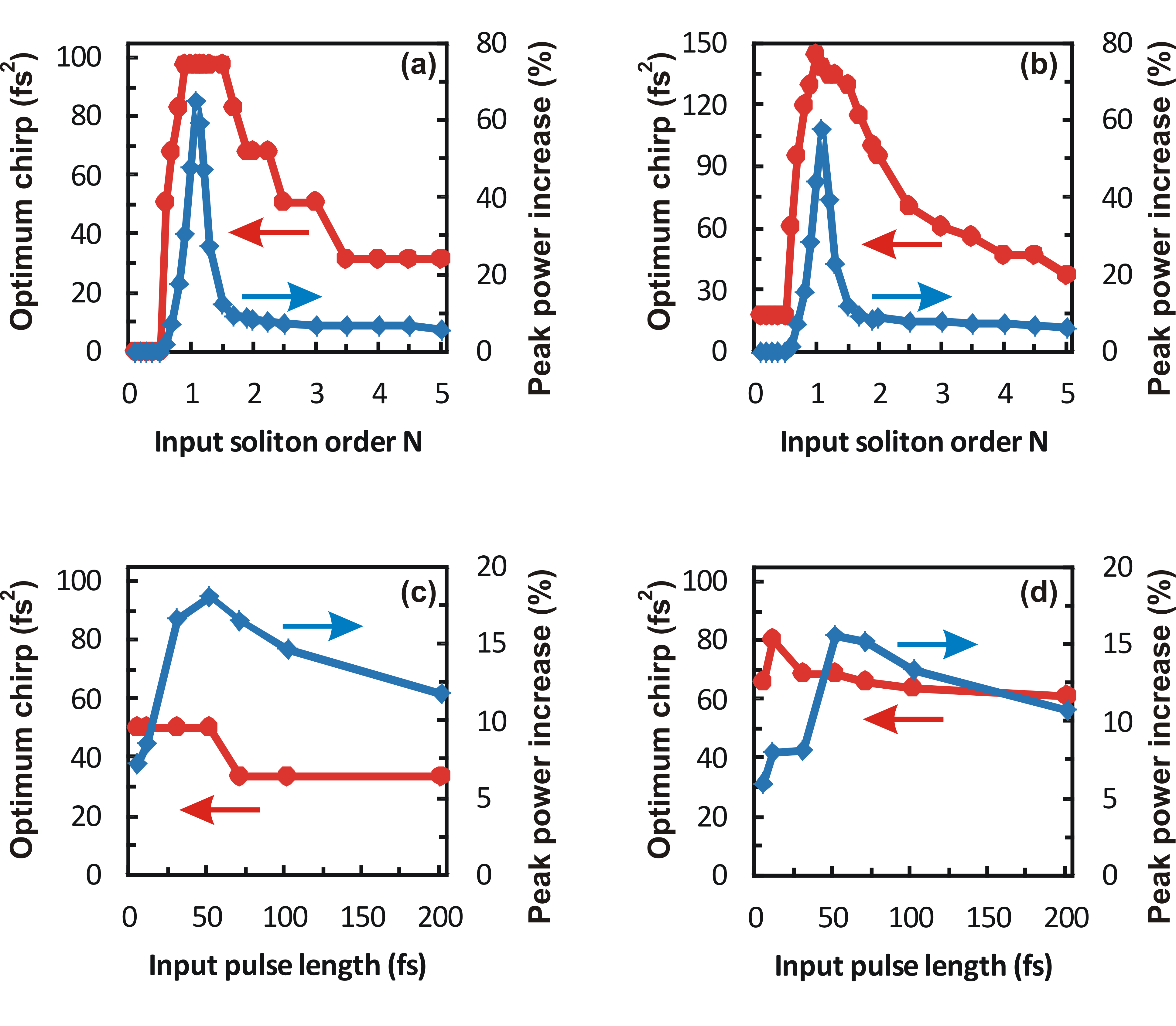}  
\end{centering}
\caption{Simulations of pulse propagation: Optimum chirp as a function of (a,b): soliton order N ,(c,d): input pulse
length for  (a,c): fiber 1, (b,d): fiber 2.\label{fig:4} }
\end{figure}

For both fibers we carried out the same simulations for $\mathrm{N}$ values up to $\mathrm{N}=5$ and a pulse lengths from $12\,\mathrm{fs}$ to $200\,\mathrm{fs}$. In all cases we found an optimum chirp for pulse compression. In Fig.\ref{fig:4} the variation in the value of this optimum chirp with soliton order and pulse length can be seen. In Fig.\ref{fig:4}(a) and in Fig.\ref{fig:4}(b) a $12\,\mathrm{fs}$ pulse length is used in each case and in Fig.\ref{fig:4}(c) and Fig.\ref{fig:4}(d) $\mathrm{N}=2.25$ is used for each pulse length. The figure shows that a positive initial chirp leads to optimal pulse compression. The obtained compressed peak powers increase between 10 and 75 percent. The chirp plays an important role for a fundamental soliton, although the pulse compression is small. For higher energies the pulse peak power increses less, but due to the strong compression RR is generated. The RR amplitude and efficiency critically depend on the pulse peak power and so we expect a systematic and strong dependence of RR on pulse chirp. The peak power enhancement applies to a large range of pulse lengths.

We also carried out simulations without self-steepening and Raman scattering. Whilst they had an impact on the position of compression in the fiber they did not significantly affect the optimum prechirp.

\section{Experiment}
 
The simulations demonstrate how the input pulse chirp can be used to optimise pulse compression.  In the experiments we investigate the impact of this on the generation efficiency of RR. We can additionally use the chirp varying technique described in \cite{McLenaghan:2014} to qualitatively investigate the evolution of RR.  

In the experiments we use short pulses centered at 800$\,$nm with a bandwidth of over 200$\,$nm from a mode-locked Ti:Sapphire laser (Rainbow, FemtoSource GmBH). These are coupled into the two photonic crystal fibers (PCFs).  The PCFs are typically around L=5$\,$mm in length as the pulses evolve over yet shorter distances. The input pulse lies in the anomalous dispersion region of the PCFs which, in the absence of higher order effects, allows for soliton formation. The PCF output in the visible and IR is measured using a CCD spectrometer (AvaSpec, Avantes BV). 

The variation in input pulse length and chirp is achieved by inserting glass elements in the beam path. Negative dispersion can be added by increasing the number of reflections from a pair of double chirped mirrors (DCMs). The minimum pulse length (FWHM) used is $7\mathrm{fs}$ for the unchirped pulse. Additional data is taken with a $12\mathrm{fs}$ pulse. The typical average input power was $100\mathrm{mW}$. The coupling into the PCFs is around $25\%$ but varies between fiber samples. 
The pulse chirp $C$ leads to an increase in the pulse length and a variation in the instantaneous frequency across the
pulse. It is given by $C=\sum_{e}\left(\left|\beta_{2}\right|_{e}z_{e}\right)+D_{\mathrm{2 DCM}}$, where $\left(\beta_{2}\right)_{e}$ and $z_{e}$ are the GVD and path distances through each dispersive element $e$ before the fiber and $D_{\mathrm{2 DCM}}$ is the negative group delay dispersion (GDD) of the DCMs.

\begin{figure}
\begin{flushleft}(a)\\
\end{flushleft}
\vspace{-1.5cm}
\begin{centering}
\quad \includegraphics[scale=0.25]{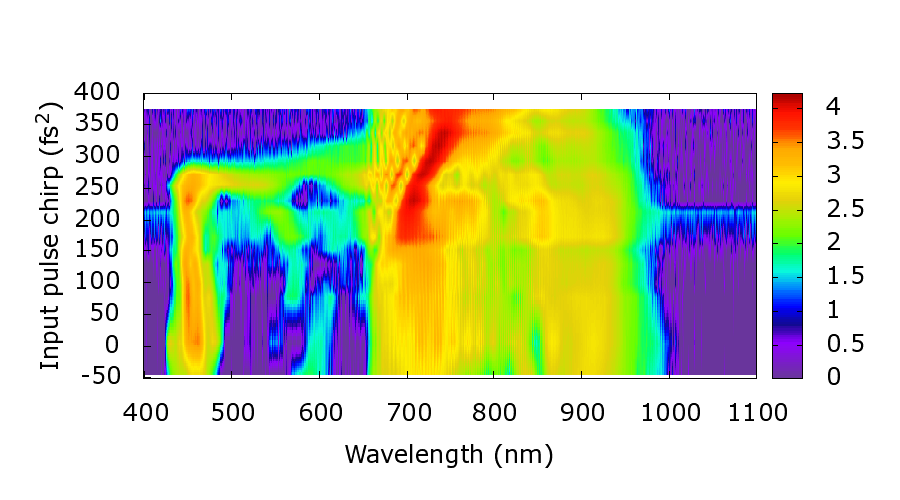}  
\end{centering}
\\
\begin{flushleft}(b)\\
\end{flushleft}
\vspace{-1.5cm}
\begin{centering}
 \quad \includegraphics[scale=0.25]{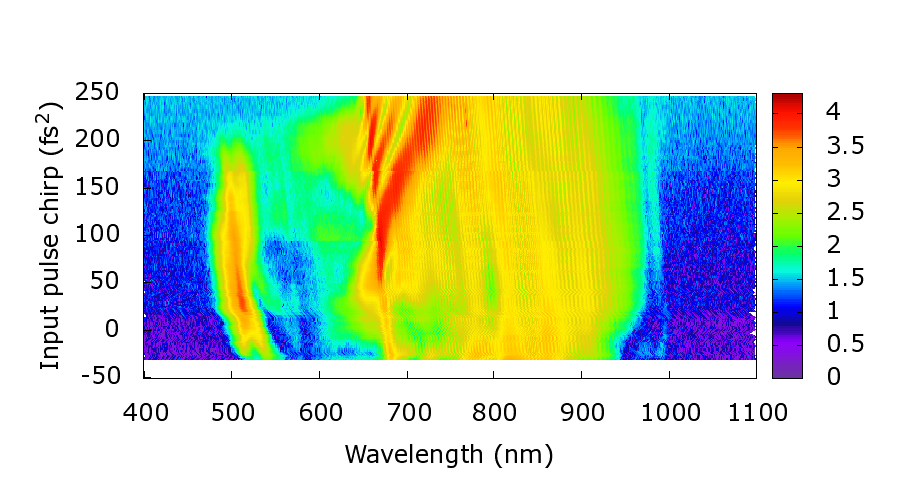}
\end{centering}
\caption{Spectral density of the visible and IR spectrum as the input chirp is
varied for (a) fiber 1 ($L$=$8\,\mathrm{mm}$, $N$=$2.25$, $\tau$=$12\,\mathrm{fs}$), (b) fiber 2 ($L$=$5\, \mathrm{mm}$, $N$=$2.11$, $\tau$=$7\,\mathrm{fs}$).\label{fig:5}}
\end{figure}

The contour plots in Fig. \ref{fig:5} show how the pulse and RR spectra vary as the input chirp is changed for the two fibers used in the experiments. RR peaks are measured at about $480\mathrm{nm}$ (fiber 1) and $510\mathrm{nm}$ (fiber 2).
The plots are qualitatively similar to those in Fig. \ref{fig:2} showing spectral evolution as a function of propagation distance. However, in this case an increasing pulse chirp corresponds to compression further along the fiber and hence a decreasing propagation distance for the RR generated at the compression point. 
For large pulse chirps the GVD is unable to compensate the chirp and the pulse does not compress fully in the fiber. Hence the output spectrum is narrow and the RR negligible. As the chirp decreases the pulse is able to fully compress at the end of the fiber and we observe the spectral broadening of the pulse and the initial generation of the RR in the fiber output. Further decrease of the chirp moves the compression point towards the input end of the fiber. The spectral broadening reduces as the pulse has temporally expanded again by the time it reaches the output. The distance propagated by the RR increases and as in the simulation plots we see it forming an isolated peak at a fixed wavelength. 

\begin{figure}
\includegraphics[scale=0.39]{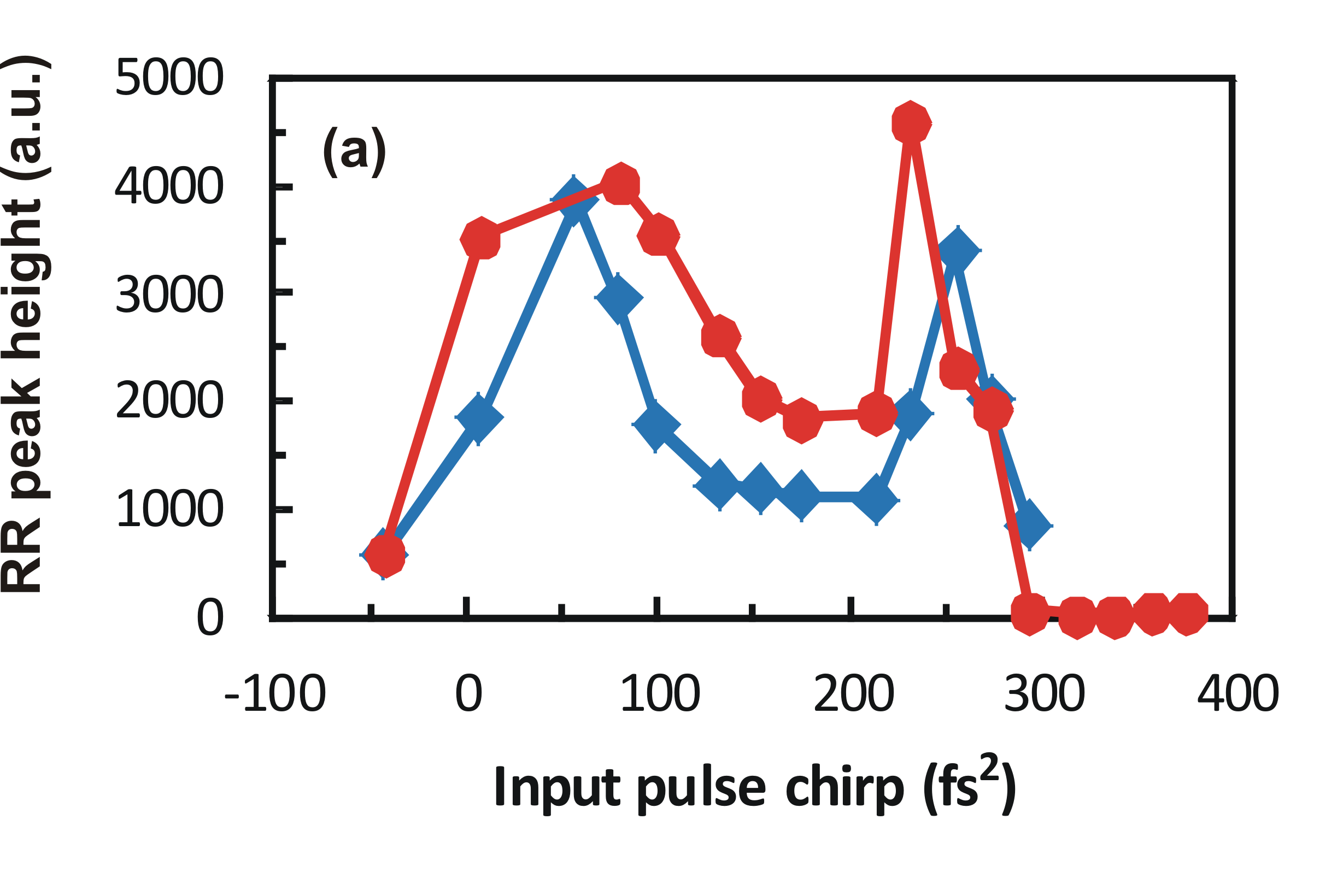}  
\quad
 \includegraphics[scale=0.39]{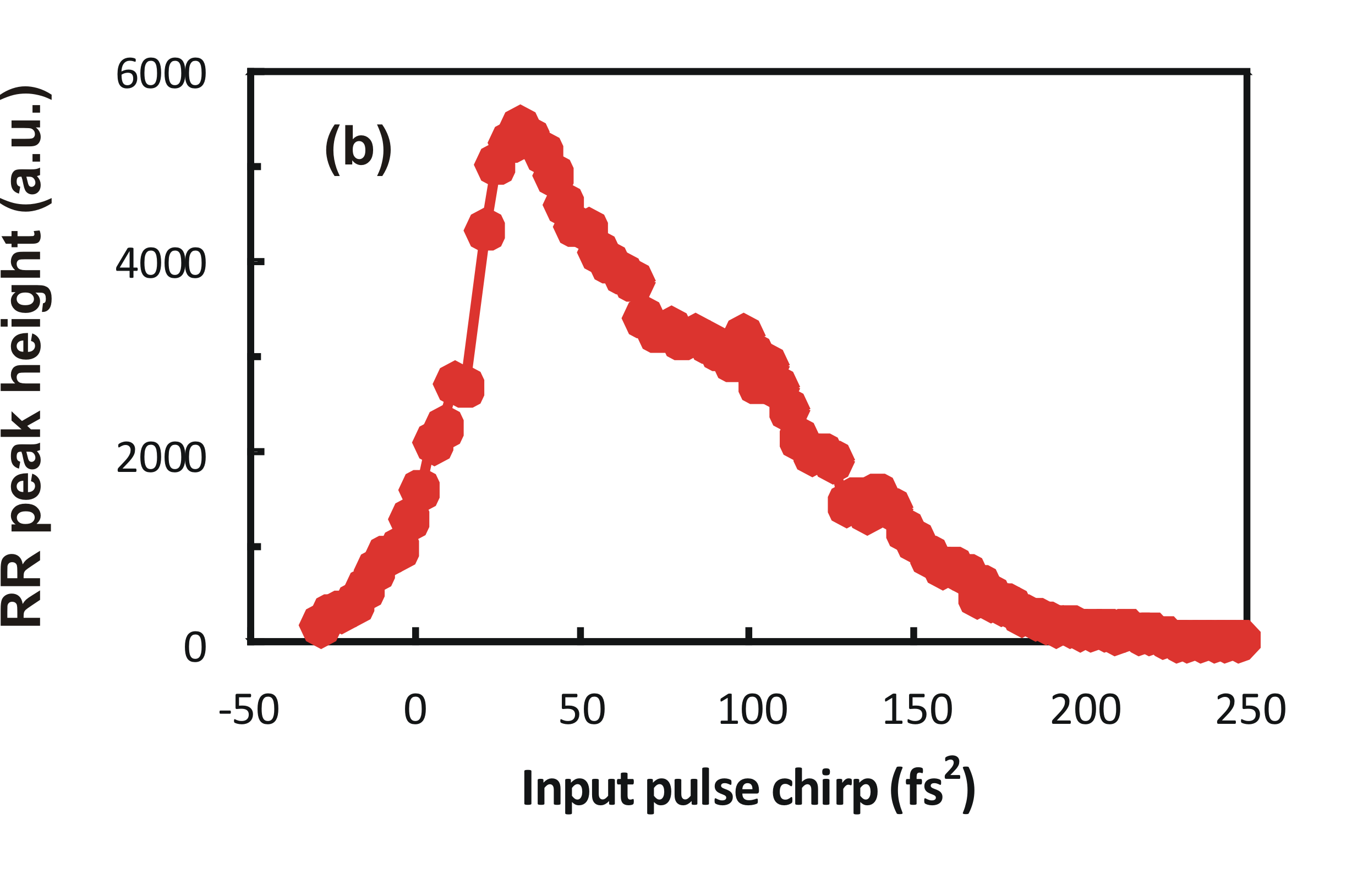}
\caption{Resonant radiation signal strength as a function of the input pulse chirp. (a)
fiber 1 ($L$=$8\,\mathrm{mm}/10\,\mathrm{mm}$ (red circles/ blue diamonds), $N$=$2.25$, $\tau$=$12\,\mathrm{fs}$), (b) fiber 2 ($L$=$5\, \mathrm{mm}$, $N$=$2.11$, $\tau$=$7\,\mathrm{fs}$).\label{fig:6}}
\end{figure}

Having seen that we can use the chirp to investigate spectral evolution we now look at the impact of chirp on the generation efficiency of RR. We expect the RR to be generated only during pulse compression and to propagate after this with little change in energy. Therefore, the total amount of RR generated should not depend on where the pulse compresses. However, in the simulations the compressed peak power slightly varied along the fiber and therefore the generation efficiency of the RR varied, too. We expect a maximum in the RR output for the optimum chirp followed by a decline as the chirp increases further until the RR becomes negligible when the chirp is too large to be compensated inside the fiber. 

The graphs in Fig. \ref{fig:6} show the RR peak height as a function of input chirp for the two fibers. In both cases the RR has the expected maximum for a small positive chirp and then declines. In Fig. \ref{fig:6}(a) the RR reaches a maximum around $50\,\mathrm{fs}^{2}$ for the two fiber lengths. This is in agreement with the simulation in Fig.  \ref{fig:4} (a). In Fig. \ref{fig:6}(b) the maximum is at around $31\,\mathrm{fs}^{2}$.  The RR peak height at the optimum chirp is over 3 times higher than at zero chirp. Note that the optimum chirp is lower than expected from   Fig. \ref{fig:4} (b,d). The difference may be due to inaccuracies in the dispersion profile of fiber 2 for IR wavelengths, which weigh much stronger if the group velocity dispersion is lower due to the larger ZDW.

Fig. \ref{fig:6} also displays a second efficient chirp corresponding to the end of the fiber; in Fig. \ref{fig:6}(a) at  $231\,\mathrm{fs}^{2}$ ($L$=$8\,\mathrm{mm}$) and  $256\,\mathrm{fs}^{2}$($L$=$10\,\mathrm{mm}$), in  Fig. \ref{fig:6}(b) at  $100\,\mathrm{fs}^{2}$.
If the compression occurs at the end of the fiber, the pulse spectrum will have expanded to overlap with the RR hence increasing the signal at that wavelength and producing the second maximum. The differences in the fiber dispersion and the pulse parameters in each case may explain why the peak is less pronounced for fiber 2. The second peak, therefore, includes light from the pulse. Otherwise, note how the RR generation follows the development of the pulse peak power in Fig. \ref{fig:3}.

In conclusion, the generation efficiency of RR can be maximized by using a small positive chirp. Such considerations are of particular importance when high conversion efficiencies are required for practical applications. Even a small improvement in the pulse compression can lead to a large increase of the RR. 
In addition, we have shown that the evolution of the pulse and RR spectra can be investigated without cutting the fiber by using the input chirp. If dispersion mircomanaged fibers, such as tapered fibers, are used, the chirp will tune this RR pulse source through the visible range.

\vspace{0.0cm}

\acknowledgements{We would like to acknowledge useful discussions with F. Biancalana and S. Kehr.}\\

\bibliographystyle{plain}

\begin{thebibliography}{10}%
\makeatletter
\providecommand \@ifxundefined [1]{%
 \@ifx{#1\undefined}
}%
\providecommand \@ifnum [1]{%
 \ifnum #1\expandafter \@firstoftwo
 \else \expandafter \@secondoftwo
 \fi
}%
\providecommand \@ifx [1]{%
 \ifx #1\expandafter \@firstoftwo
 \else \expandafter \@secondoftwo
 \fi
}%
\providecommand \natexlab [1]{#1}%
\providecommand \enquote  [1]{``#1''}%
\providecommand \bibnamefont  [1]{#1}%
\providecommand \bibfnamefont [1]{#1}%
\providecommand \citenamefont [1]{#1}%
\providecommand \href@noop [0]{\@secondoftwo}%
\providecommand \href [0]{\begingroup \@sanitize@url \@href}%
\providecommand \@href[1]{\@@startlink{#1}\@@href}%
\providecommand \@@href[1]{\endgroup#1\@@endlink}%
\providecommand \@sanitize@url [0]{\catcode `\\12\catcode `\$12\catcode
  `\&12\catcode `\#12\catcode `\^12\catcode `\_12\catcode `\%12\relax}%
\providecommand \@@startlink[1]{}%
\providecommand \@@endlink[0]{}%
\providecommand \url  [0]{\begingroup\@sanitize@url \@url }%
\providecommand \@url [1]{\endgroup\@href {#1}{\urlprefix }}%
\providecommand \urlprefix  [0]{URL }%
\providecommand \Eprint [0]{\href }%
\providecommand \doibase [0]{http://dx.doi.org/}%
\providecommand \selectlanguage [0]{\@gobble}%
\providecommand \bibinfo  [0]{\@secondoftwo}%
\providecommand \bibfield  [0]{\@secondoftwo}%
\providecommand \translation [1]{[#1]}%
\providecommand \BibitemOpen [0]{}%
\providecommand \bibitemStop [0]{}%
\providecommand \bibitemNoStop [0]{.\EOS\space}%
\providecommand \EOS [0]{\spacefactor3000\relax}%
\providecommand \BibitemShut  [1]{\csname bibitem#1\endcsname}%
\let\auto@bib@innerbib\@empty
\bibitem{Agrawal:01}
Govind~P. Agrawal.
\newblock {\em Nonlinear Fiber Optics}.
\newblock Optics and Photonics. Academic Press, 3rd edition, 2001.

\bibitem{Akhmediev:1995hc}
Karlsson~Magnus Akhmediev, Nail.
\newblock Cherenkov radiation emitted by solitons in optical fibers.
\newblock {\em Physical Review A}, 51(3):2602--2607, 03 1995.

\bibitem{Chang:10}
Guoqing Chang, Li-Jin Chen, and Franz~X. K\"{a}rtner.
\newblock Highly efficient cherenkov radiation in photonic crystal fibers for
  broadband visible wavelength generation.
\newblock {\em Opt. Lett.}, 35(14):2361--2363, Jul 2010.

\bibitem{Chang:12}
Guoqing Chang, Chih-Hao Li, A.~Glenday, G.~Furesz, N.~Langellier, Li-Jin Chen,
  M.W. Webber, Jinkang Lim, Hung-Wen Chen, D.F. Phillips, A.~Szentgyorgyi, R.L.
  Walsworth, and F.X. Kartner.
\newblock Spectrally flat, broadband visible-wavelength astro-comb.
\newblock In {\em Lasers and Electro-Optics (CLEO), 2012 Conference on}, pages
  1--2, 2012.

\bibitem{Cheng2011}
Chunfu Cheng, Youqing Wang, and Qinghua Lv.
\newblock Effect of initial frequency chirp on the supercontinuum generation in
  all-normal dispersion photonic crystal fibers.
\newblock In {\em Photonics and Optoelectronics Meetings (POEM) 2011: Optical
  Communication Systems and Networking}, volume 8331, pages 83310O--83310O--8,
  2011.

\bibitem{Choudhary:12}
Amol Choudhary and Friedrich K\"{o}nig.
\newblock Efficient frequency shifting of dispersive waves at solitons.
\newblock {\em Opt. Express}, 20(5):5538--5546, Feb 2012.

\bibitem{Cristiani:04}
Ilaria Cristiani, Riccardo Tediosi, Luca Tartara, and Vittorio Degiorgio.
\newblock Dispersive wave generation by solitons in microstructured optical
  fibers.
\newblock {\em Opt. Express}, 12(1):124--135, Jan 2004.

\bibitem{Belgiorno:10}
M.Clerici V.Gorini G.Ortenzi L. Rizzi E. Rubino V.G.Sala D.~Faccio F.Belgiorno,
  S.L.Cacciatori.
\newblock Hawking radiation from ultrashort laser pulse filaments.
\newblock {\em Phys Rev Lett}, 105(20), November 2010.

\bibitem{Fu2004}
Xiquan Fu, Liejia Qian, Shuangchun Wen, and Dianyuan Fan.
\newblock Nonlinear chirped pulse propagation and supercontinuum generation in
  microstructured optical fibre.
\newblock {\em Journal of Optics A: Pure and Applied Optics}, 6(11):1012--,
  2004.

\bibitem{Herrmann:2002fj}
J.~Herrmann, U.~Griebner, N.~Zhavoronkov, A.~Husakou, D.~Nickel, J.~C. Knight,
  W.~J. Wadsworth, P.~St.~J. Russell, and G.~Korn.
\newblock Experimental evid2ence for supercontinuum generation by fission of
  higher-order solitons in photonic fibers.
\newblock {\em Physical Review Letters}, 88(17):173901--, 04 2002.

\bibitem{Husakou:2001rt}
A~V Husakou and J~Herrmann.
\newblock Supercontinuum generation of higher-order solitons by fission in
  photonic crystal fibers.
\newblock {\em Phys Rev Lett}, 87(20):203901, Nov 2001.

\bibitem{Ivanov:06}
Anatoly~A. Ivanov, Mikhail~V. Alfimov, Aleksei~M. Zheltikov, Marcin Szpulak,
  Waclaw Urbanczyk, and Jan W\'{o}jcik.
\newblock Polarization-controlled vectorial spectral transformations of
  femtosecond pulses in a birefringent photonic-crystal fiber.
\newblock {\em J. Opt. Soc. Am. B}, 23(5):986--991, May 2006.

\bibitem{Liu:13}
X.~Liu, G.E. Villanueva, J.~Laegsgaard, U.~Moller, H.~Tu, S.A. Boppart, and
  D.~Turchinovich.
\newblock Low-noise operation of all-fiber femtosecond cherenkov laser.
\newblock {\em Photonics Technology Letters, IEEE}, 25(9):892--895, May 2013.

\bibitem{Liu:12}
Xiaomin Liu, Jesper L{\ae}gsgaard, Uffe M{\o}ller, Haohua Tu, Stephen~A.
  Boppart, and Dmitry Turchinovich.
\newblock All-fiber femtosecond cherenkov radiation source.
\newblock {\em Opt. Lett.}, 37(13):2769--2771, Jul 2012.

\bibitem{Lu:05}
Fei Lu, Yujun Deng, Larry~E. Foulkrod, and Wayne~H. Knox.
\newblock Dispersion micro-managed holey fiber and coherent blue-violet
  continuum generation.
\newblock In {\em Conference on Lasers and Electro-Optics/Quantum Electronics
  and Laser Science and Photonic Applications Systems Technologies}, page CWA2.
  Optical Society of America, 2005.

\bibitem{Lu:06}
Fei Lu and Wayne~H. Knox.
\newblock Generation, characterization, and application of broadband coherent
  femtosecond visible pulses in dispersion micromanaged holey fibers.
\newblock {\em J. Opt. Soc. Am. B}, 23(6):1221--1227, Jun 2006.

\bibitem{McLenaghan:2014}
J.~McLenaghan and F.~Koenig.
\newblock {Few-cycle fiber pulse compression and evolution of negative resonant
  radiation}.
\newblock {\em {New J. Phys.}}, {16}, {JUN 10} {2014}.

\bibitem{Mitrofanov:06}
Aleksandr~V. Mitrofanov, Yaroslav~M. Linik, Ryszard Buczynski, Dariusz Pysz,
  Dusan Lorenc, Ignac Bugar, Anatoly~A. Ivanov, Mikhail~V. Alfimov, Andrei~B.
  Fedotov, and Aleksei~M. Zheltikov.
\newblock Highly birefringent silicate glass photonic-crystal fiber with
  polarization-controlled frequency-shifted output: A promising fiber light
  source fornonlinear raman microspectroscopy.
\newblock {\em Opt. Express}, 14(22):10645--10651, Oct 2006.

\bibitem{Paschotta}
R.~Paschotta.
\newblock {\em simulation software PROPULSE}.
\newblock RP Photonics Consulting GmbH, Zurich, Switzerland.

\bibitem{Philbin:2008fr}
Thomas~G. Philbin, Chris Kuklewicz, Scott Robertson, Stephen Hill, Friedrich
  K{\"o}nig, and Ulf Leonhardt.
\newblock Fiber-optical analog of the event horizon.
\newblock {\em Science}, 319(5868):1367--1370, 03 2008.

\bibitem{Roy:2009kl}
Samudra Roy, S.~K. Bhadra, and Govind~P. Agrawal.
\newblock Dispersive waves emitted by solitons perturbed by third-order
  dispersion inside optical fibers.
\newblock {\em Physical Review A}, 79(2):023824--, 02 2009.

\bibitem{Rubino:2012fk}
E.~Rubino, A.~Lotti, F.~Belgiorno, S.~L. Cacciatori, A.~Couairon, U.~Leonhardt,
  and D.~Faccio.
\newblock Soliton-induced relativistic-scattering and amplification.
\newblock {\em Sci. Rep.}, 2, 12 2012.

\bibitem{Rubino:2012ly}
E.~Rubino, J.~McLenaghan, S.~C. Kehr, F.~Belgiorno, D.~Townsend, S.~Rohr, C.~E.
  Kuklewicz, U.~Leonhardt, F.~K{\"o}nig, and D.~Faccio.
\newblock Negative-frequency resonant radiation.
\newblock {\em Physical Review Letters}, 108(25):253901--, 06 2012.

\bibitem{Skryabin:2003qy}
D.~V. Skryabin, F.~Luan, J.~C. Knight, and P.~St.~J. Russell.
\newblock Soliton self-frequency shift cancellation in photonic crystal fibers.
\newblock {\em Science}, 301(5640):1705--1708, 09 2003.

\bibitem{Tianprateep2004}
M.~Tianprateep, Ji~Tada, T.~Yamazaki, and F.~Kannari.
\newblock Spectral-shape-controllable supercontinuum generation in
  microstructured fibers using adaptive pulse shaping technique.
\newblock {\em Jpn. J. Appl. Phys.}, 43:8059--8063, 2004.

\bibitem{Tianprateep2005}
Montian Tianprateep, Junji Tada, and Fumihiko Kannari.
\newblock Influence of polarization and pulse shape of femtosecond initial
  laser pulses on spectral broadening in microstructure fibers.
\newblock {\em Optical Review}, 12(3):179--189--, 2005.

\bibitem{Tu:09}
Haohua Tu and Stephen~A. Boppart.
\newblock Optical frequency up-conversion by supercontinuum-free widely-tunable
  fiber-optic cherenkov radiation.
\newblock {\em Opt. Express}, 17(12):9858--9872, Jun 2009.

\bibitem{Tu:09b}
Haohua Tu and Stephen~A. Boppart.
\newblock Ultraviolet-visible non-supercontinuum ultrafast source enabled by
  switching single silicon strand-like photonic crystal fibers.
\newblock {\em Opt. Express}, 17(20):17983--17988, Sep 2009.

\bibitem{Tu:2013}
Haohua Tu and Stephen~A. Boppart.
\newblock {Coherent fiber supercontinuum for biophotonics}.
\newblock {\em {Laser Photon. Rev.}}, {7}({5}):{628--645}, {SEP} {2013}.

\bibitem{Zhang2007}
Hua Zhang, Song Yu, Jie Zhang, and Wanyi Gu.
\newblock Effect of frequency chirp on supercontinuum generation in photonic
  crystal fibers with two zero-dispersion wavelengths.
\newblock {\em Opt. Express}, 15(3):1147--1154, 2007.

\bibitem{Zhu2004}
Zhaoming Zhu and Thomas Brown.
\newblock Effect of frequency chirping on supercontinuum generation in photonic
  crystal fibers.
\newblock {\em Opt. Express}, 12(4):689--694, 2004.
\end{thebibliography}
%

\end{document}